\documentclass[%
 reprint,
 amsmath,amssymb,
 aps,
prb,
nolongbibliography,
]{revtex4-2}

\usepackage{graphicx}
\usepackage{dcolumn}
\usepackage{bm}
\usepackage{xspace}
\usepackage{setspace}
\usepackage{subfig}
\usepackage{verbatim}
\usepackage{booktabs} 
\usepackage{siunitx}
\usepackage{amsmath}
\usepackage{ragged2e}
\usepackage{soul}
\usepackage{afterpage}

\usepackage[
    colorlinks=true,   
    linkcolor=blue,     
    citecolor=blue,     
    urlcolor=blue,      
    bookmarks=true,     
    bookmarksopen=true  
]{hyperref}

\begin{document}

\title{Interplay between Quantum Metric and Hybridized Collective Mode in Flat-Band Superfluids}

\author{Yi Liu}
\email{lyfsf5121@gmail.com}

\author{Mingyan Wang}

\author{Penghui Hu}

\author{Yao Lu}
\email{yao.y.lu1203@gmail.com}

\affiliation{College of Information Science and Engineering, Huaqiao University, Xiamen 361021, China} 

\date{\today}

\begin{abstract}
We investigate collective excitations in flat-band superfluids by incorporating the coupled dynamics of pairing (phase and amplitude) and density fluctuations. We demonstrate that for any time-reversal symmetric superfluid system with an isolated flat band, only a single low-energy collective mode emerges in the long-wavelength limit. In contrast to the linearly dispersive Goldstone mode in conventional superfluids, this hybridized mode is gapless at zero momentum but exhibits a quadratic dispersion ($\omega \propto q^2$) at small momenta. We show that the dispersion coefficients of this collective mode are governed by the normal-state quantum metric of the flat band. These analytical predictions are in excellent agreement with numerical calculations. Our results are applicable to any generic $s$-wave flat-band superfluid, provided the flat band is energetically well separated from other dispersive bands.
\end{abstract}

\maketitle

\section{Introduction}
\label{sec:intro}

\begin{figure}[htbp]
    \centering
    \includegraphics[width=0.70\linewidth]{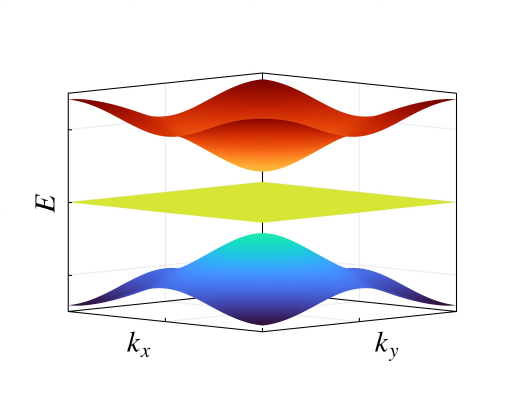} 
    \caption{Schematic illustration of the normal-state band structure for a prototypical isolated flat-band system.}
    \label{fig:flat_band_schematic}
\end{figure}

A flat band is a dispersionless energy band where the single-particle kinetic energy is quenched, leading to a diverging density of states. As shown in Fig.~\ref{fig:flat_band_schematic}, the spectrum features an isolated flat band situated between two dispersive bands. Due to the absence of kinetic energy, even arbitrarily weak electron-electron interactions can dominate the system's physics, paving the way for a rich variety of strongly correlated quantum phases. One of the most fascinating phenomena in these systems is unconventional superconductivity. The experimental discovery of correlated insulator states and superconductivity in magic-angle twisted bilayer graphene (MATBG) \cite{cao2018correlated, cao2018unconventional, cao2020strange, cao2021nematicity, yankowitz2019tuning, lu2019superconductors, jiang2019charge} has ignited a significant surge of both experimental and theoretical interest in flat-band superfluids and superconductors.

In conventional Bardeen--Cooper--Schrieffer (BCS) theory, the superfluid weight is inversely proportional to the effective mass of the paired electrons \cite{tinkham2004introduction}. For a strictly isolated flat band, the effective mass diverges, which implies a vanishing superfluid weight. However, recent theoretical advances have revealed that in flat-band superfluids, the superfluid weight is bounded from below by the quantum geometric properties of the Bloch wavefunctions, specifically the normal-state quantum metric \cite{peotta2015superfluidity, liang2017band, torma2018quantum, torma2022superconductivity, julku2020superfluid, julku2021quantum, huhtinen2022revisiting, tian2023evidence}. This established that the quantum metric replaces the conventional kinetic energy as the primary origin of superfluidity in flat bands. 

Beyond static properties, dynamic collective excitations—namely, the phase (Goldstone) and amplitude (Higgs) modes \cite{nambu1960quasi, goldstone1961field, goldstone1962broken, anderson1963plasmons, littlewood1981gauge, littlewood1982amplitude, matsunaga2013higgs, anderson1958random, volkov1974collisionless, pekker2015amplitude, shimano2020higgs, iskin2024cooper}—provide fundamental insights into the stability and low-energy response of the superconducting state. Extending these concepts to flat-band systems, recent studies have systematically investigated the behavior of these collective modes \cite{lewandowski2019intrinsically, kuang2021collective, stauber2016quasi, kauppila2016collective, xiao2024quantum, wu2021superfluid, julku2021excitations, tovmasyan2016effective, villegas2021anomalous}. For instance, by utilizing Hubbard--Stratonovich decouplings in a dice lattice, it has been shown that the gapless Goldstone mode maintains a conventional linear acoustic dispersion relation ($\omega \propto q$), with its sound velocity determined by the quantum metric in the flat-band limit \cite{wu2021superfluid}.

However, existing theoretical treatments predominantly rely on single-channel approximations, thereby neglecting the dynamic influence of particle density fluctuations on the collective modes. In flat-band superfluids, the bandwidth is much smaller than the interaction energy scale, establishing the hierarchy $W \ll \Delta$, where $W$ is the isolated flat-band width and $\Delta$ is the pairing gap. In this regime, variations in the particle density profoundly influence the superfluid properties, causing the superconducting phase and the low-energy particle density fluctuations to become intrinsically intertwined in the long-wavelength limit. 

In this work, we formulate a multi-channel field-theoretical framework to systematically resolve the collective excitations in generic flat-band superfluids. By employing the Generalized Hubbard--Stratonovich (GHS) transformation, we treat the pairing fluctuations (phase and amplitude) and the particle density fluctuations on an equal footing.

\section{Model}
\label{sec:model}

We consider a generic two-dimensional lattice model hosting an isolated flat band, which is partially filled. In the flat-band limit, the single-particle kinetic energy is quenched, rendering the degeneracy highly susceptible to inter-particle interactions. The system can be described by the following microscopic Hamiltonian:
\begin{equation}
    H = H_0 + H_{\text{int}},
\end{equation}
where $H_0$ is the non-interacting Hamiltonian, and $H_{\text{int}}$ represents the attractive interaction. The single-particle part is described by the generic non-interacting lattice Hamiltonian:
\begin{equation}
    H_0 = \sum_{\bm{r},\bm{r}'} \sum_{\sigma} \bm{c}_{\sigma}^\dagger(\bm{r}) \left[ h_0(\bm{r} - \bm{r}') - \mu \delta_{\bm{r},\bm{r}'} I \right] \bm{c}_{\sigma}(\bm{r}'),
\end{equation}
where $\bm{c}_{\sigma}(\bm{r}) = [c_{1\sigma}(\bm{r}), c_{2\sigma}(\bm{r}), \dots, c_{N_b\sigma}(\bm{r})]^T$ denotes the $N_b$-component annihilation operator vector in the unit cell $\bm{r}$. Here, $c_{\gamma\sigma}(\bm{r})$ annihilates a fermion at unit cell position $\bm{r}$, sublattice $\gamma$ ($\gamma = 1, 2, \dots, N_b$), with spin $\sigma \in \{\uparrow, \downarrow\}$. The $N_b \times N_b$ matrix $h_0(\bm{r} - \bm{r}')$ represents the real-space hopping matrix between different unit cells. $I$ is the $N_b \times N_b$ identity matrix, and $\mu$ is the chemical potential. The single-particle band structure features a flat band with constant energy $\epsilon_0$, separated from other dispersive bands by a finite energy gap.

We consider an on-site attractive Hubbard interaction, which drives the $s$-wave superfluidity:
\begin{equation}
    H_{\text{int}} = -U \sum_{\bm{r},\gamma} n_{\gamma\uparrow}(\bm{r}) n_{\gamma\downarrow}(\bm{r}).
\end{equation}
Here, $U > 0$ denotes the strength of the local attractive interaction. The operator $n_{\gamma\sigma}(\bm{r}) = c_{\gamma\sigma}^\dagger(\bm{r}) c_{\gamma\sigma}(\bm{r})$ is the particle number operator at unit cell position $\bm{r}$ and sublattice $\gamma$. To study the superfluid state, we apply the mean-field approximation. We define the sublattice-dependent, real $s$-wave order parameters as $\Delta_\gamma = -U \langle c_{\gamma\downarrow}(\bm{r}) c_{\gamma\uparrow}(\bm{r}) \rangle$, which preserves the full structural generality of the lattice geometry. In terms of the real-space Nambu spinor $\Psi(\bm{r}) = (\bm{c}_{\uparrow}(\bm{r}), \bm{c}_{\downarrow}^\dagger(\bm{r}))^T$, the mean-field Hamiltonian is given by:
\begin{equation}
    H_{\text{MF}} = \sum_{\bm{r}, \bm{r}'} \Psi^\dagger(\bm{r}) \mathcal{H}_{\text{BdG}}(\bm{r} - \bm{r}') \Psi(\bm{r}'),
\end{equation}
where the $2N_b \times 2N_b$ real-space BdG matrix is:
\begin{equation}
    \mathcal{H}_{\text{BdG}}(\bm{r} - \bm{r}') = 
    \begin{pmatrix}
        \xi(\bm{r} - \bm{r}') & \hat{\Delta}(\bm{r} - \bm{r}') \\
        \hat{\Delta}^\dagger(\bm{r} - \bm{r}') & -\xi^*(\bm{r} - \bm{r}')
    \end{pmatrix}.
\end{equation}
Here, $\xi(\bm{r} - \bm{r}') = h_0(\bm{r} - \bm{r}') - \mu \delta_{\bm{r},\bm{r}'} I$ defines the single-particle energy relative to the chemical potential, and $\hat{\Delta}(\bm{r} - \bm{r}') = \delta_{\bm{r},\bm{r}'} \text{diag}(\Delta_1, \Delta_2, \dots, \Delta_{N_b})$ represents the localized, sublattice-dependent pairing matrix.

Provided the chemical potential $\mu$ is near the flat band and the interaction strength $U$ is smaller than the energy gap separating the flat band from the dispersive bands, the low-energy physics is governed by the flat-band subspace. By setting the flat-band energy to zero ($\epsilon_0 = 0$) and projecting the BdG Hamiltonian onto the flat band, the effective real-space Hamiltonian can be generally expressed as a tensor product of the Nambu matrix and the flat-band spatial projector:
\begin{equation}
    \label{eq:Heff}
    \mathcal{H}_{\text{eff}}(\bm{r}, \bm{r}^{\prime}) = (-\mu \tau_z + \Delta_0 \tau_x) \otimes \mathcal{P}(\bm{r}, \bm{r}^{\prime}),
\end{equation}
where $\mathcal{P}(\bm{r}, \bm{r}^{\prime}) = \sum_{n \in \text{FB}} \psi_n(\bm{r}) \psi_n^\dagger(\bm{r}^{\prime})$ is the projection operator onto the isolated flat-band subspace spanned by the single-particle basis wavefunctions $\{\psi_n\}$, and $\tau_{x,y,z}$ are the Pauli matrices acting on the particle-hole space. Crucially, the effective flat-band pairing gap is defined as $\Delta_0 = \sum_{\gamma=1}^{N_b} \Delta_\gamma |\psi_{n,\gamma}|^2$, where $\psi_{n,\gamma}$ represents the intra-cell wavefunction amplitude on sublattice $\gamma$. Under the uniform pairing condition, this density-weighted summation is independent of the state index $n$, mapping the microscopic sublattice-dependent order parameters onto a robust restricted subspace.

The corresponding quasiparticle excitation spectrum is strictly flat with a single-particle excitation energy $E = \sqrt{\mu^2 + \Delta_0^2}$. To study the collective excitations, we must proceed beyond the mean-field level to incorporate the Gaussian fluctuations of the order parameter and particle density via the path-integral formalism. Focusing on the low-energy acoustic modes in the long-wavelength limit $ |\bm{q}| \ll 1/a$, where $a$ is the lattice constant, we assume that the amplitude, phase, and density fluctuations synchronize across all sublattices, behaving as uniform fields. The effective Hamiltonian $\mathcal{H}_{\text{eff}}$ provides the mathematical foundation to analytically resolve these multi-channel dynamics within the restricted Hilbert space. Throughout the subsequent derivations, the system is evaluated in the zero-temperature limit $T=0$.

\section{Analytical Formalism}
\label{sec:analytical}

\subsection{Generalized Hubbard--Stratonovich Transformation}

To resolve the collective excitations and go beyond the mean-field approximation, we formulate the problem within the path-integral framework. Starting from the microscopic Hamiltonian $H = H_0 + H_{\text{int}}$ introduced in Sec.~\ref{sec:model}, the grand canonical partition function is defined as the thermal trace over the many-body Fock space:
\begin{equation} \label{eq:trace_Z}
    \mathcal{Z} = \mathrm{Tr} e^{-\beta H},
\end{equation}
where $\beta = 1/T$ is the inverse temperature, and the chemical potential $\mu$ is implicitly included in $H_0$. Throughout this work, we adopt natural units by setting $k_{\mathrm{B}} = \hbar = 1$.

To evaluate this trace and handle the non-commutativity between $H_0$ and $H_{\text{int}}$, we employ the Suzuki-Trotter decomposition. By introducing the continuous imaginary time $\tau = it$ via the standard Wick rotation, the physical time domain is mapped to the thermal interval $\tau \in [0, \beta]$. Dividing this imaginary time interval into $N_\tau$ discrete slices of infinitesimal length $\epsilon = \beta/N_\tau$, indexed by the integer $l \in \{1, 2, \dots, N_\tau\}$, the partition function is expressed in the time-sliced operator representation:
\begin{equation} \label{eq:time_sliced_Z}
    \mathcal{Z} = \lim_{\epsilon \to 0} \mathrm{Tr} \left\{ T_\tau \prod_{l=1}^{N_\tau} \left[ 1 - \epsilon H_{0,l} - \epsilon H_{\text{int},l} \right] \right\},
\end{equation}
where $T_\tau$ is the imaginary time ordering operator, and the subscript $l$ labels the operators evaluated at the discrete imaginary time $\tau_l = l\epsilon$. Here, the time-sliced interaction Hamiltonian is expressed in terms of the discrete local four-fermion operator as $H_{\text{int},l} = -U \sum_{\bm{r}, \gamma} \hat{\Omega}_{l, \gamma}(\bm{r})$, where $\hat{\Omega}_{l, \gamma}(\bm{r}) = c_{\gamma\uparrow,l}^\dagger(\bm{r}) c_{\gamma\downarrow,l}^\dagger(\bm{r}) c_{\gamma\downarrow,l}(\bm{r}) c_{\gamma\uparrow,l}(\bm{r})$ represents the on-site interaction density at unit cell $\bm{r}$ and sublattice $\gamma$.

However, a direct application of the standard Hubbard--Stratonovich (HS) transformation fails to capture the quantum fluctuations in both the density (particle-hole) and pairing (particle-particle) channels simultaneously. To resolve this and describe the multi-channel dynamics of flat-band systems on an equal footing, we adopt the Generalized Hubbard--Stratonovich (GHS) transformation, following the foundational formulation by Kerman et al \cite{joglekar2001microscopic, kerman1981mean, kerman1983functional, kerman1984functional}. We apply the Generalized Hubbard--Stratonovich (GHS) transformation directly to the discrete time-sliced operator representation. 

This multi-channel decomposition yields the partition function:
\begin{equation}
    \mathcal{Z} = \lim_{\epsilon \to 0} \int \prod_{l=1}^{N_\tau} \mathcal{D}[\phi, \Delta, \bar{\Delta}] e^{-S[\phi, \Delta, \bar{\Delta}]},
\end{equation}

\begin{align} \label{eq:GHS_action}
    S[\phi, \Delta, \bar{\Delta}] &= \epsilon \sum_{l=1}^{N_\tau} \int d^2\bm{r} \left[ \frac{\phi_{l}^2(\bm{r})}{2U} + \frac{|\Delta_{l}(\bm{r})|^2}{U} \right] \nonumber \\
    &\quad - \log \mathrm{Tr} \left\{ T_\tau \prod_{l=1}^{N_\tau} \left[ 1 - \epsilon \tilde{H}_{M,l} - \epsilon^2 \tilde{H}_{\text{res},l} \right] \right\},
\end{align}
with
\begin{equation}
    \tilde{H}_{M,l} = H_{0,l} - \tilde{H}_{\phi,l} - \tilde{H}_{\Delta,l},
\end{equation}
\begin{equation}
    \tilde{H}_{\phi,l} = \sum_{\gamma} \int d^2\bm{r} \frac{1}{\sqrt{2}} \phi_{l, \gamma}(\bm{r}) \sum_{\sigma} c_{\gamma\sigma,l}^\dagger(\bm{r}) c_{\gamma\sigma,l}(\bm{r}),
\end{equation}
\begin{equation}
    \tilde{H}_{\Delta,l} = \sum_{\gamma} \int d^2\bm{r} \left[ \Delta_{l, \gamma}(\bm{r}) c_{\gamma\uparrow,l}^\dagger(\bm{r}) c_{\gamma\downarrow,l}^\dagger(\bm{r}) + \mathrm{H.c.} \right],
\end{equation}
\begin{equation}
\begin{split}
    \tilde{H}_{\text{res},l} &= \sum_{\gamma} \int d^2\bm{r} \frac{U}{c_1+c_2} \hat{\Omega}_{l, \gamma}(\bm{r}) \\
    &\quad \times \left[ c_1\frac{\phi_{l, \gamma}^2(\bm{r})}{V \mathcal{N}^2} + c_2\frac{|\Delta_{l, \gamma}(\bm{r})|^2}{V \mathcal{N}^2} \right],
\end{split}
\end{equation}
where $V$ is the strength of the trial interaction \cite{joglekar2001microscopic, kerman1981mean, kerman1983functional, kerman1984functional}. Here $V$ is set to be $V = U$. $\mathcal{N}^2$ stands for a repeat sum over the $\mathcal{N}$ single-particle state labels in the Hilbert space. $c_1$ and $c_2$ are two positive real numbers, and the results are independent of the choices of $c_1$ and $c_2$. For simplicity, we set $c_1=1$, $c_2=2$.

If we ignore the path integral over bosonic fields, the interaction term is in high order of $\epsilon$, which can be dropped, then we recover the mean-field approximation in Sec.~\ref{sec:model}. The mean-field potential $\{\Delta_0, \bar{\Delta}_0, \phi_0\}$ are given by the minimization of the mean-field action $S_0$ \cite{joglekar2001microscopic}. Going beyond the mean-field theory, we restore the path integral and write the bosonic fields as $\Delta_{l}(\bm{r}) \approx \Delta_0 + \rho_{l}(\bm{r}) - i\Delta_0 \theta_{l}(\bm{r})$, $\bar{\Delta}_{l}(\bm{r}) \approx \Delta_0 + \rho_{l}(\bm{r}) + i\Delta_0 \theta_{l}(\bm{r})$, $\phi_{l}(\bm{r}) = \phi_0 + \delta\phi_{l}(\bm{r})$. Expanding the action around its minimum up to the second order of $\rho$, $\theta$ and $\delta\phi$, we obtain: 
\begin{equation}
    S \approx S_0 + \frac{1}{2} \sum_{l,l'} \int d^2\bm{r} d^2\bm{r}' \bar{\Phi}_l(\bm{r}) \mathcal{M}_{l,l'}(\bm{r},\bm{r}') \Phi_{l'}(\bm{r}'), 
\end{equation}
where the bosonic fields are defined by $\Phi(\bm{r}) = (\rho, \Delta_0 \theta, \delta\phi)^T$, $\bar{\Phi}(\bm{r}) = (\rho, \Delta_0 \theta, \delta\phi)$. For convenience, we also define the bosonic operators $\hat{\Phi}(\bm{r}) = (\hat{\rho}, \Delta_0 \hat{\theta}, \delta\hat{\phi})^T$, with their forms given by:
\begin{align}
    \hat{\rho}(\bm{r}) &= \sum_{\gamma} \left[ c_{\gamma\uparrow}^\dagger(\bm{r}) c_{\gamma\downarrow}^\dagger(\bm{r}) + c_{\gamma\downarrow}(\bm{r}) c_{\gamma\uparrow}(\bm{r}) \right], \\
    \Delta_0 \hat{\theta}(\bm{r}) &= i \sum_{\gamma} \left[ c_{\gamma\downarrow}(\bm{r}) c_{\gamma\uparrow}(\bm{r}) - c_{\gamma\uparrow}^\dagger(\bm{r}) c_{\gamma\downarrow}^\dagger(\bm{r}) \right], \\
    \delta\hat{\phi}(\bm{r}) &= \frac{1}{\sqrt{2}} \sum_{\gamma, \sigma} c_{\gamma\sigma}^\dagger(\bm{r}) c_{\gamma\sigma}(\bm{r}).
\end{align}

The matrix $\mathcal{M}$ is given by: 
\begin{equation}
\begin{split}
    &\mathcal{M}_{l,l'}^{ij}(\bm{r},\bm{r}') = \frac{\partial^2 S}{\partial \bar{\Phi}_l^{i}(\bm{r}) \partial \Phi_{l'}^j(\bm{r}')} \\
    &= \epsilon \big[ \delta_{l,l'} U^{-1} + \epsilon(1-\delta_{l,l'}) D_{l,l'} + \epsilon\delta_{l,l'} \mathcal{S} \big].
\end{split}
\end{equation}
Here only time indices are explicitly shown. The matrices $D$ and $S$ are defined by: 
\begin{equation}
\begin{split}
    D_{l,l'}^{ij}(\bm{r},\bm{r}') &= \langle T_\tau \hat{\Phi}_l^i(\bm{r}) \hat{\Phi}_{l'}^{j\dagger}(\bm{r}') \rangle_0 - \langle \hat{\Phi}_l^i(\bm{r}) \rangle_0 \langle \hat{\Phi}_{l'}^{j\dagger}(\bm{r}') \rangle_0,
\end{split}
\end{equation}
\begin{equation}
\begin{split}
    \mathcal{S}_{l,l'}^{ij}(\bm{r},\bm{r}') &= \frac{\delta_{ij}\delta(\bm{r}-\bm{r}') V \sum_{\gamma} \int d^2\tilde{\bm{r}} \langle \hat{\Omega}_{l, \gamma}(\tilde{\bm{r}}) \rangle}{3U\mathcal{N}^2}  \\
    &\quad + \langle \hat{\Phi}_l^i(\bm{r}) \rangle_0 \langle \hat{\Phi}_{l'}^{j\dagger}(\bm{r}') \rangle_0,
\end{split}
\end{equation}
where $\langle \dots \rangle_0$ denotes the mean-field thermodynamic average. As explained in Ref.~\cite{kerman1983functional, kerman1984functional}, the $\mathcal{S}$ matrix represents the contribution of the single quasi-particle motion which remains beyond the mean-field grand potential, and thus does not contribute to the dynamics of the collective excitations.

\subsection{The Multi-Channel Dynamic Response Matrix}
\label{subsec:response_matrix}

To extract the dispersion relation $\omega(\bm{q})$ of the collective excitation modes, we must solve the equation $\det[U^{-1}\bm{I} + \bm{D}] = 0$. 

To analytically resolve this equation, we transform the relevant real-space matrices into the momentum and Matsubara frequency spaces. For convenience, we define the full $3 \times 3$ dynamic response matrix as:
\begin{equation} \label{eq:M_matrix_full}
    \bm{M}(\bm{q}, i\omega_m) = \bm{M}_{\text{bare}} + \bm{D}(\bm{q}, i\omega_m).
\end{equation}

The first term is the bare coupling matrix, $\bm{M}_{\text{bare}} = \text{diag} \left( 1/U, 1/U, 1/U \right)$. The second term, $\bm{D}(\bm{q}, i\omega_m)$, is the dynamic polarization matrix corresponding to the Fourier transform of the spatio-temporal correlator $D_{l,l'}^{ij}(\bm{r}, \bm{r}')$. 

By expressing the fluctuation operators $\hat{\Phi}$ in the Nambu spinor basis and applying Wick's theorem, the evaluation of the four-fermion expectation values factorizes into products of single-particle flat-band Green's functions $\mathcal{G}_0(\bm{k}, i\omega_n)$. Consequently, the dynamic matrix elements, now indexed by the Nambu pseudo-spin components $\alpha, \beta \in \{x,y,z\}$, evaluate to the compact fermionic trace:
\begin{align} \label{eq:bubble_elements}
    D_{\alpha\beta}(\bm{q}, i\omega_m) &= \frac{1}{2\beta} \sum_{\bm{k}, i\omega_n} \mathrm{Tr} \Big[ \tau_\alpha \mathcal{G}_0(\bm{k}+\bm{q}, i\omega_n+i\omega_m) \nonumber \\
    &\quad \quad \quad \quad \quad \quad \times \tau_\beta \mathcal{G}_0(\bm{k}, i\omega_n) \Big].
\end{align}

By projecting $\mathcal{G}_0$ onto the Bogoliubov quasiparticle eigenbasis and evaluating the internal fermionic Matsubara summation $\omega_n = (2n+1)\pi/\beta$, followed by the analytic continuation $i\omega_m \rightarrow \omega + i0^+$, the dynamic matrix elements cleanly separate into resonant and anti-resonant transition channels:
\begin{equation} \label{eq:bubble_T1_T2}
    D_{\alpha\beta}(\bm{q}, \omega) = -\frac{1}{2} \sum_{\bm{k}} \left[ \frac{T_1(\bm{k}, \bm{q})}{2E - \omega - i0^+} + \frac{T_2(\bm{k}, \bm{q})}{2E + \omega + i0^+} \right],
\end{equation}
where $E = \sqrt{\mu^2 + \Delta_0^2}$. The terms $T_1(\bm{k}, \bm{q})$ and $T_2(\bm{k}, \bm{q})$ are defined as:
\begin{align}
    T_1(\bm{k}, \bm{q}) &= t_\alpha^{34} t_\beta^{43} \left| \langle \psi_{\bm{k}} | \psi_{\bm{k}+\bm{q}} \rangle \right|^2, \\
    T_2(\bm{k}, \bm{q}) &= t_\alpha^{43} t_\beta^{34} \left| \langle \psi_{\bm{k}} | \psi_{\bm{k}+\bm{q}} \rangle \right|^2 = T_1^*(\bm{k}, \bm{q}).
\end{align}
Here, $t_\alpha^{s's} = \langle \phi_{s'} | \tau_\alpha | \phi_s \rangle$ represents the matrix element of the Pauli matrix $\tau_\alpha$ evaluated within the isolated flat-band subspace, where $|\phi_s\rangle$ denotes the Nambu eigenstate corresponding to the band index $s \in \{3, 4\}$. The detailed evaluation of the Matsubara summation, along with the explicit analytical forms of these matrix elements, are provided in Appendix~\ref{app:matsubara_evaluation}.

Solving for the dispersion relation $\omega(\bm{q})$ is thus mathematically equivalent to finding the roots of the determinant equation:
\begin{equation} \label{eq:secular_det}
    \det \left[ \bm{M}(\bm{q}, \omega) \right] = 0.
\end{equation}

By systematically expanding this secular determinant in the long-wavelength limit, we will analytically extract the explicit low-energy collective modes in the following sections.

\subsection{Zero-Frequency Pole}
\label{subsec:omega0}

The collective mode dispersion $\omega(\bm{q})$ is determined by the roots of the secular equation $\det[\bm{M}(\bm{q}, \omega)] = 0$. In the long-wavelength limit ($|\bm{q}| \ll 1/a$), we expand $\omega(\bm{q})$ with respect to the momentum $\bm{q} = (q_x, q_y)^T$:
\begin{equation} \label{eq:omega_q_expansion}
    \omega(\bm{q}) = \omega_0 + \bm{v} \cdot \bm{q} + \bm{q}^T \bm{\kappa} \bm{q} + \mathcal{O}(q^3).
\end{equation}
Here, $\omega_0 \equiv \omega(\bm{0})$, $\bm{v}$ denotes the sound velocity vector, and $\bm{\kappa}$ is the $2 \times 2$ quadratic dispersion matrix.

To determine $\omega_0$, we evaluate the secular equation at $\bm{q}=\bm{0}$, $\det[\bm{M}(0, \omega_0)] = 0$, analytically by enforcing the mean-field gap equation constraint. As detailed in Appendix~\ref{app:det_derivation}, this yields: 
\begin{equation}
    \det[\bm{M}(0, \omega_0)] = \frac{-N_k^3 \omega_0^2}{8E^3 (4E^2 - \omega_0^2)},
\end{equation}
where $N_k$ is the number of unit cells and $E = \sqrt{\mu^2 + \Delta_0^2}$. Equating this determinant to zero identifies $\omega_0 = 0$ as the unique physical root.

\subsection{Vanishing of the Linear Term}
\label{subsec:vanishing_linear}

Since the collective mode satisfies $\omega_0=0$, the dispersion relation reduces to $\omega(\bm{q}) = \bm{v} \cdot \bm{q} + \mathcal{O}(q^2)$ in the long-wavelength limit ($q \equiv |\bm{q}| \ll 1/a$). Substituting this expansion into the response matrix, we can expand $\bm{M} \left[ \bm{q}, \omega(\bm{q}) \right]$ up to linear order in the momentum vector components:
\begin{equation} \label{eq:matrix_taylor_first_order}
\begin{split}
    \bm{M} \left[ \bm{q}, \omega(\bm{q}) \right] & = \bm{M}(0,0) + (\bm{v} \cdot \bm{q}) \left. \frac{\partial \bm{M}}{\partial \omega} \right|_{(0,0)} \\
    & \quad + \left. (\bm{q} \cdot \nabla_{\bm{q}}) \bm{M} \right|_{(0,0)} + \mathcal{O}(q^2).
\end{split}
\end{equation}

To evaluate the first-order momentum gradient in Eq.~(\ref{eq:matrix_taylor_first_order}), we inspect the geometric origin of the dynamic polarization bubble $\bm{D}(\bm{q}, \omega)$. Under the flat-band constraint where the single-particle dispersion is quenched to a constant value ($\varepsilon_{\bm{k}} = \varepsilon_0$), all momentum dependence is mapped onto the normal-state squared wave-function overlap $F(\bm{k}, \bm{q}) = |\langle \psi_{\bm{k}} | \psi_{\bm{k}+\bm{q}} \rangle|^2$. Performing a gradient expansion on the shifted Bloch state yields:
\begin{equation} \label{eq:bloch_linear_gradient}
    |\psi_{\bm{k}+\bm{q}}\rangle = |\psi_{\bm{k}}\rangle + (\bm{q} \cdot \nabla_{\bm{k}}) |\psi_{\bm{k}}\rangle + \mathcal{O}(q^2).
\end{equation}

The inner product with the unshifted state gives the overlap $\langle \psi_{\bm{k}} | \psi_{\bm{k}+\bm{q}}\rangle = 1 + \bm{q} \cdot \langle \psi_{\bm{k}} | \nabla_{\bm{k}} \psi_{\bm{k}}\rangle + \mathcal{O}(q^2)$. Multiplying by its complex conjugate yields the squared overlap to linear order:
\begin{equation} \label{eq:overlap_linear_expanded}
    F(\bm{k}, \bm{q}) = 1 + \bm{q} \cdot \left( \langle \nabla_{\bm{k}} \psi_{\bm{k}} | \psi_{\bm{k}}\rangle + \langle \psi_{\bm{k}} | \nabla_{\bm{k}} \psi_{\bm{k}}\rangle \right) + \mathcal{O}(q^2).
\end{equation}

Because the Bloch states are inherently normalized ($\langle \psi_{\bm{k}} | \psi_{\bm{k}}\rangle = 1$), taking the momentum gradient of this normalization condition yields:
\begin{equation} \label{eq:normalization_derivative_identity}
    \nabla_{\bm{k}} \langle \psi_{\bm{k}} | \psi_{\bm{k}}\rangle = \langle \nabla_{\bm{k}} \psi_{\bm{k}} | \psi_{\bm{k}}\rangle + \langle \psi_{\bm{k}} | \nabla_{\bm{k}} \psi_{\bm{k}}\rangle \equiv \bm{0}.
\end{equation}

Equation~(\ref{eq:normalization_derivative_identity}) demonstrates that the linear cross-term in the wave-function overlap expansion vanishes at each wavevector, which implies that the first-order momentum gradient of the response matrix is zero:
\begin{equation} \label{eq:macroscopic_momentum_derivative_zero}
    \left. \nabla_{\bm{q}} \bm{M} \right|_{(0,0)} = \bm{0}.
\end{equation}

Substituting Eq.~(\ref{eq:macroscopic_momentum_derivative_zero}) into Eq.~(\ref{eq:matrix_taylor_first_order}), the linearized expansion is driven solely by the dynamic frequency derivative. To evaluate this surviving linear term, we first examine the unperturbed matrix $\bm{M}(0,0)$. As explicitly derived in Appendix~\ref{app:null_space_derivation}, $\bm{M}(0,0)$ contains a two-dimensional null space spanned by the pure phase fluctuation basis $|v_1\rangle = (0, 1, 0)^T$ and the amplitude-density hybridized basis $|v_2\rangle = (\mu/E, 0, \Delta_0/E)^T$.

To proceed, we partition the full response matrix $\bm{M}$ into the low-energy null space (denoted by index $L$) and its high-energy orthogonal complement (denoted by index $H$), yielding the block structure $\bm{M} = \begin{pmatrix} \bm{M}_{LL} & \bm{M}_{LH} \\ \bm{M}_{HL} & \bm{M}_{HH} \end{pmatrix}$. By definition, the cross-couplings vanish at $\bm{q}=\bm{0}$ and $\omega=0$ ($\bm{M}_{LH}(0,0) = \bm{0}$). Since the first-order spatial gradient identically vanishes (Eq.~\ref{eq:macroscopic_momentum_derivative_zero}), the cross-couplings are bounded by $\mathcal{O}(q)$. 

As justified by the Schur determinant identity detailed in Appendix~\ref{app:schur_justification}, the feedback correction from the massive high-energy space takes the form $-\bm{M}_{LH} \bm{M}_{HH}^{-1} \bm{M}_{HL}$. Because $\bm{M}_{HH}^{-1}$ is non-singular ($\mathcal{O}(1)$), this entire correction term is of order $\mathcal{O}(q) \cdot \mathcal{O}(1) \cdot \mathcal{O}(q) = \mathcal{O}(q^2)$. Consequently, up to linear order in momentum, solving the full secular equation $\det[\bm{M}]=0$ is equivalent to solving the reduced projected equation $\det[\bm{M}_{LL}]=0$. Therefore, to determine the sound velocity vector $\bm{v}$, we project the first-order frequency expansion of the response matrix directly onto the degenerate subspace $\mathcal{P} = \text{span}\{|v_1\rangle, |v_2\rangle\}$ to construct the effective $2 \times 2$ matrix $\bm{M}_{LL}$.

Because the unperturbed matrix elements vanish identically upon projection ($\langle v_i | \bm{M}(0,0) | v_j \rangle = 0$), $\bm{M}_{LL}$ is given by:
\begin{equation} \label{eq:low_energy_M_first_order}
    \left[ \bm{M}_{LL}(\bm{q}, \omega) \right]_{ij} = (\bm{v} \cdot \bm{q}) \bm{A}_{ij} + \mathcal{O}(q^2),
\end{equation}
where $\bm{A}$ represents the frequency derivative matrix projected onto the null space, with elements defined by $\bm{A}_{ij} = \langle v_i | \left. \frac{\partial \bm{M}}{\partial \omega} \right|_{(0,0)} | v_j \rangle$. The analytical evaluation of this projected matrix via the Nambu spinor transition matrix elements (detailed in Appendix~\ref{app:matrix_A_evaluation}) yields the skew-Hermitian block:
\begin{equation} \label{eq:matrix_A_explicit}
    \bm{A} = \frac{N_k}{4E^2} \begin{pmatrix} 0 & i \\ -i & 0 \end{pmatrix}.
\end{equation}

Substituting $\bm{A}$ into Eq.~(\ref{eq:low_energy_M_first_order}), the linear-order projected secular equation becomes $\det[(\bm{v} \cdot \bm{q}) \bm{A}] = 0$. Because the matrix $\bm{A}$ is manifestly non-singular ($\det \bm{A} = ( \frac{N_k}{4E^2} )^2 \neq 0$) and the scattering momentum is finite ($q \neq 0$), this algebraic constraint forces the inner product to vanish:
\begin{equation} \label{eq:final_velocity_zero}
    \bm{v} \cdot \bm{q} = 0.
\end{equation}

Since this condition must hold for an arbitrary momentum direction $\bm{q}$, the sound velocity vector must strictly vanish ($\bm{v} = \bm{0}$). This root confirms that the low-energy collective excitation mode in a flat-band superfluid lacks a linear acoustic term. Consequently, the dispersion relation Eq.~(\ref{eq:omega_q_expansion}) simplifies to $\omega(\bm{q}) = \bm{q}^T \bm{\kappa} \bm{q} + \mathcal{O}(q^3)$, enabling us to proceed to the next perturbation order to determine the geometric origin of the quadratic dispersion tensor $\bm{\kappa}$.

\subsection{The Geometric Origin of the Quadratic Dispersion}
\label{subsec:schur_quadratic_dispersion_coefficient}

With the sound velocity vector vanishing ($\bm{v} = \bm{0}$), the gapless collective mode dispersion relation simplifies to $\omega(\bm{q}) = \bm{q}^T \bm{\kappa} \bm{q} + \mathcal{O}(q^3)$. To determine the quadratic dispersion tensor $\bm{\kappa}$, we expand the response matrix $\bm{M} \left[ \bm{q}, \omega(\bm{q}) \right]$ to second order around $(\bm{0},0)$:
\begin{equation} \label{eq:matrix_taylor_second_order}
\begin{split}
    \bm{M} \left[ \bm{q}, \omega(\bm{q}) \right] &= \bm{M}(0,0) + (\bm{q}^T \bm{\kappa} \bm{q}) \left. \frac{\partial \bm{M}}{\partial \omega} \right|_{(0,0)} \\
    &\quad + \frac{1}{2} \left. (\bm{q} \cdot \nabla_{\bm{q}})^2 \bm{M} \right|_{(0,0)} + \mathcal{O}(q^3),
\end{split}
\end{equation}
where $\nabla_{\bm{q}}$ is the momentum gradient operator. We apply the Schur complement technique to project this fluctuation matrix onto the degenerate subspace $\mathcal{P} \equiv \text{span}\{|v_1\rangle, |v_2\rangle\}$. Since the sound velocity vanishes ($\bm{v} = \bm{0}$) [Eq.~(\ref{eq:final_velocity_zero})] and the first-order momentum gradient is zero [Eq.~(\ref{eq:macroscopic_momentum_derivative_zero})], the cross-coupling block $\bm{M}_{LH}$ is of order $\mathcal{O}(q^2)$. Consequently, the high-energy feedback correction $-\bm{M}_{LH} \bm{M}_{HH}^{-1} \bm{M}_{HL}$ is of order $\mathcal{O}(q^4)$ and can be safely discarded up to second order in momentum.

Upon projection, the unperturbed matrix elements vanish identically ($\langle v_i | \bm{M}(0,0) | v_j \rangle = 0$). The linear frequency derivative block maps directly onto the matrix $\bm{A}$ defined in Eq.~(\ref{eq:matrix_A_explicit}), where $\bm{A}_{ij} = \langle v_i | \left. \frac{\partial \bm{M}}{\partial \omega} \right|_{(0,0)} | v_j \rangle$. Concurrently, by expanding the spatial gradient operator into its Cartesian components $a, b \in \{x, y\}$, we define a set of matrices $\bm{B}_{ab}$: 
\begin{equation} \label{eq:matrix_B_definition}
    \left[ \bm{B}_{ab} \right]_{ij} = \frac{1}{2} \langle v_i | \left. \frac{\partial^2 \bm{M}}{\partial q_a \partial q_b} \right|_{(0,0)} | v_j \rangle.
\end{equation}
As detailed in Appendix~\ref{app:matrix_B_evaluation}, the microscopic evaluation reveals that $\bm{B}_{ab}$ is proportional to the identity matrix within the subspace, taking the explicit form $\left[ \bm{B}_{ab} \right]_{ij} = \bar{B}_{ab} \delta_{ij}$, with the scalar coefficient defined by the normal-state quantum metric elements: $\bar{B}_{ab} = \frac{1}{2E} \sum_{\bm{k}} g_{ab}(\bm{k})$. Here, the quantum metric $g_{ab}(\bm{k})$ is defined via the long-wavelength expansion of the squared Bloch states overlap: 
\begin{equation} \label{eq:quantum_metric_definition}
    |\langle \psi_{\bm{k}} | \psi_{\bm{k}+\bm{q}} \rangle|^2 = 1 - \sum_{a,b} g_{ab}(\bm{k}) q_a q_b + \mathcal{O}(q^3).
\end{equation}

Gathering these components and substituting the quadratic dispersion $\omega(\bm{q}) = \sum_{a,b} q_a q_b \kappa_{ab}$, the effective $2 \times 2$ long-wavelength secular equation up to order $\mathcal{O}(q^2)$ can be elegantly unified under a single momentum summation:
\begin{equation} \label{eq:low_energy_secular_second_order}
    \det \left[ \sum_{a,b} q_a q_b \Big( \kappa_{ab} \bm{A} + \bm{B}_{ab} \Big) \right] = 0.
\end{equation}

Substituting the explicit off-diagonal elements of $\bm{A}$ and the diagonal elements of $\bm{B}_{ab}$, the secular determinant expands analytically to:
\begin{equation}
    \left( \sum_{a,b} q_a q_b \bar{B}_{ab} \right)^2 - \left( \sum_{a,b} q_a q_b \kappa_{ab} \right)^2 \left( \frac{N_k}{4E^2} \right)^2 = 0.
\end{equation}
Taking the square root and isolating the dispersion term $\sum_{a,b} \kappa_{ab} q_a q_b$ yields:
\begin{equation} \label{eq:tensor_polynomial_identity}
    \sum_{a,b} \kappa_{ab} q_a q_b = \frac{4E^2}{N_k} \sum_{a,b} q_a q_b \left( \frac{1}{2E} \sum_{\bm{k}} g_{ab}(\bm{k}) \right).
\end{equation}

Because this algebraic constraint must hold identically for any arbitrary scattering momentum vector $\bm{q}$, the symmetric tensor coefficients on both sides of Eq.~(\ref{eq:tensor_polynomial_identity}) must be strictly equal. This directly yields the analytical expression for the quadratic dispersion tensor:
\begin{equation}
    \kappa_{ab} = \frac{2E}{N_k} \sum_{\bm{k}} g_{ab}(\bm{k}).
\end{equation}

This relation demonstrates that in flat-band superfluids, the quadratic dispersion tensor $\kappa_{ab}$ of the collective excitation mode is governed by the quantum metric $g_{ab}(\bm{k})$ of the normal-state flat band.

\section{Numerical Results}
\label{sec:numerical}

To verify the analytical predictions, we numerically evaluate a two-dimensional flat-band model: the Lieb lattice with staggered hoppings \cite{lieb1989two, julku2016geometric}. Throughout this section, energies and lengths are measured in units of the nearest-neighbor hopping amplitude $J$ and the lattice constant $a$, respectively (i.e., we set $J=a=1$), rendering the momentum $\bm{q}$ dimensionless.

For the numerical calculations, we set the mean-field pairing gap on the $A$ and $C$ sublattices to $\Delta_0 = 0.2J$ ($\Delta_B = 0$). To break the particle-hole symmetry and induce dynamic channel hybridization, we introduce a finite chemical potential $\mu = 0.1J$. The staggered hopping amplitude $\delta$, which controls the normal-state band gap and the quantum metric, is varied within $\delta \in [0.3J, 0.5J]$. The roots of the multi-channel secular equation $\det[\bm{M}(\bm{q}, \omega)] = 0$ are extracted numerically over a discretized Brillouin zone.

\subsection{BdG Spectrum and 3D Collective Dispersion}

\begin{figure}[htbp]
    \centering
    \includegraphics[width=0.9\linewidth]{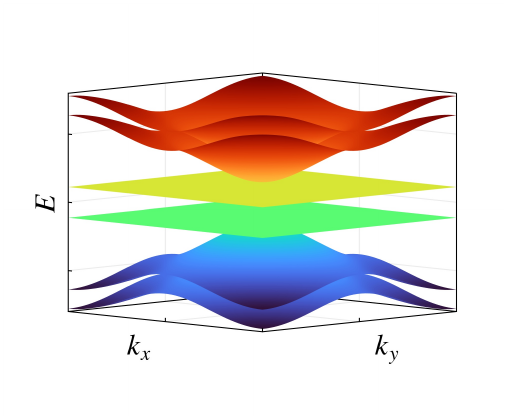} 
    \caption{Schematic diagram of the $6 \times 6$ Bogoliubov--de Gennes quasiparticle excitation spectrum for the Lieb lattice with a finite chemical potential and a staggered hopping amplitude.}
    \label{fig:band_structure}
\end{figure}

Upon introducing the uniform $s$-wave pairing field $\Delta_0$, the $6 \times 6$ Bogoliubov--de Gennes (BdG) spectrum features a pair of isolated, perfectly flat bands at $\pm E$ (where $E = \sqrt{\mu^2 + \Delta_0^2}$), separated from the adjacent dispersive bands by a finite energy gap (Fig.~\ref{fig:band_structure}). This energetic isolation justifies our low-energy effective projection.

\begin{figure}[htbp]
    \centering
    \includegraphics[width=0.85\linewidth]{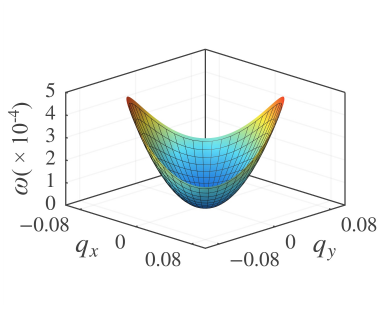} 
    \caption{The low-energy collective mode dispersion $\omega(q_x, q_y)$, obtained by numerically solving $\det[\bm{M}(\bm{q}, \omega)] = 0$ for small momenta at $\delta = 0.3J$.}
    \label{fig:3d_dispersion}
\end{figure}

By tracking the roots of $\det[\bm{M}(\bm{q}, \omega)] = 0$, we map the 3D dispersion landscape of the hybridized collective mode at small momenta (Fig.~\ref{fig:3d_dispersion}). The gapless excitation surface emerges directly from the origin ($\bm{q}=\bm{0}, \omega_0 = 0$) and exhibits a parabolic profile ($\omega \propto q^2$).

\subsection{Absence of Linear Term and Scaling Analysis}

\begin{figure}[htbp]
    \centering
    \includegraphics[width=\linewidth]{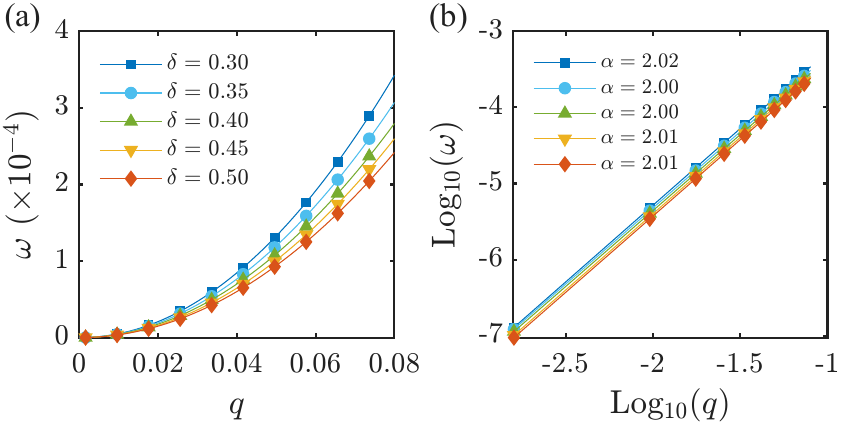} 
    \caption{Scaling analysis of the collective mode dispersion along the $x$ direction ($\bm{q} = q_x \hat{\bm{x}}$). (a) The extracted low-energy dispersion curves $\omega(q_x)$ for various $\delta$. (b) Log-log plot of the dispersion data, yielding power-law exponents of $\alpha \approx 2.0$.}
    \label{fig:scaling}
\end{figure}

To systematically verify the vanishing sound velocity under varying geometric conditions, we evaluate the dispersion trajectories along the $x$ direction ($\bm{q} = q_x \hat{\bm{x}}$, i.e., $q_y = 0$) for various staggered hopping parameters $\delta$ [Fig.~\ref{fig:scaling}(a)].

Assuming a power-law dispersion $\omega \propto q_x^\alpha$, a log-log scaling analysis on these numerical curves [Fig.~\ref{fig:scaling}(b)] consistently yields exponents of $\alpha \approx 2.0$ across all selected parameters. This confirms the parabolic nature of the hybridized collective mode and verifies the absence of a linear acoustic term, in agreement with our analytical proof.

\subsection{Verification of the Quadratic Dispersion Coefficient}

Having established the parabolic scaling behavior $\omega \simeq \kappa_{xx} q_x^2$, we proceed to verify that the quadratic dispersion coefficient $\kappa_{xx}$ is governed by the normal-state quantum metric. For each $\delta$, we calculate the Brillouin-zone-averaged quantum metric, $\frac{1}{N_k} \sum_{\bm{k}} g_{xx}(\bm{k})$. Concurrently, the dispersion curvature $\kappa_{xx}$ is extracted from a parabolic fit to the numerical roots. 

\begin{figure}[htbp]
    \centering
    \includegraphics[width=0.85\linewidth]{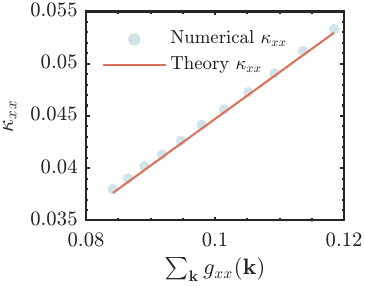} 
    \caption{Collective mode dispersion curvature $\kappa_{xx}$ versus the average normal-state quantum metric under varying staggered hoppings $\delta$. The circular markers represent numerical data from the multi-channel response matrix, while the solid line represents the analytical prediction.}
    \label{fig:metric_verification}
\end{figure}

Figure~\ref{fig:metric_verification} exhibits a direct comparison between these numerically fitted coefficients (circles) and our theoretical prediction $\kappa_{xx} = \frac{2E}{N_k} \sum_{\bm{k}} g_{xx}(\bm{k})$ (solid line). The data points closely track the theoretical linear dependence across the entire parameter regime, quantitatively confirming the geometric origin of the collective mode dispersion derived via our formalism.

\section{Conclusion and Discussion}
\label{sec:conclusion}

In summary, we have employed a multi-channel field-theoretical framework based on the Generalized Hubbard--Stratonovich (GHS) transformation to investigate low-energy collective excitations in flat-band superfluids. This framework incorporates the pairing (phase and amplitude) and particle density fluctuations simultaneously, circumventing the overcounting of microscopic interactions that arises when standard Hubbard--Stratonovich decouplings are extended to multiple channels. To extract the response in the long-wavelength limit, we projected out the massive intra-cell optical modes, reducing the microscopic degrees of freedom to a uniform three-component acoustic fluctuation field. This projection enables the formulation of the response matrix, allowing for the analytical resolution of the dynamic hybridization among the amplitude, phase, and density channels in isolated flat bands.

The collective mode dispersion relations are determined analytically by solving the secular equation $\det[\bm{M}(\bm{q}, \omega)] = 0$ within the low-energy degenerate subspace via the Schur complement technique. First, within the GHS framework, incorporating the particle density channel into the flat-band superfluid system yields a vanishing sound velocity, $\bm{v} = \bm{0}$. Second, we obtain an analytical expression for the quadratic dispersion tensor, $\kappa_{ab} = \frac{2E}{N_k} \sum_{\bm{k}} g_{ab}(\bm{k})$, which demonstrates that the dispersion relation is governed by the normal-state quantum metric of the Bloch states. This further establishes that the collective excitation mode is dominated by the quadratic dispersion when both channels are simultaneously taken into account.

The analytical expression for $\kappa_{ab}$ reveals that the quadratic dispersion is independent of the chemical potential $\mu$. The algebraic contributions from the finite chemical potential are completely absorbed into the total quasiparticle energy gap through the flat-band identity $\mu^2 + \Delta_0^2 = E^2$. Consequently, the collective mode dispersion remains unaffected by the specific filling fraction, provided the system remains well within the flat-band projection regime where dispersive bands are energetically excluded. Furthermore, the low-temperature evaluations show no explicit temperature dependence, indicating that the framework can be extended to finite temperatures where thermal fluctuations are well below the pairing gap ($T \ll E$).

Previously, the GHS framework has been successfully employed to implement dual-channel decouplings in flat-band-like systems, yielding consistent results in specific regimes. For instance, in the lowest Landau levels, avoiding microscopic overcounting requires the simultaneous incorporation of both direct (density) and exchange (pseudospin) channels, yielding a similar quadratic collective dispersion \cite{joglekar2001microscopic}. Furthermore, fluctuation analyses in pseudo-magnetic flat bands show that coupled fluctuations in the Hartree and pairing channels lead to a $q^2$ dispersion governed by the geometric length scale of the pseudo-Landau levels \cite{li2019orbital}. In the present work, we simultaneously incorporate the superconducting pairing and particle density channels in a generic flat-band superfluid, analytically deriving the explicit expression for the quadratic dispersion tensor.

These analytical findings are supported by numerical calculations on a Lieb lattice with staggered hoppings. The calculated three-dimensional dispersion spectrum exhibits a gapless parabolic dispersion ($\omega \propto q^2$) emerging from the origin. The quantitative agreement between the numerically fitted dispersion coefficients and the analytical predictions across a range of staggered hoppings confirms the geometric origin of the collective mode dispersion.

\begin{acknowledgments}
     Y. L. acknowledges the support from Huaqiao University (605-50Y24031) and the support from Natural Science Foundation of Xiamen China (605-52424128).
\end{acknowledgments}

\bibliography{references}

\onecolumngrid
\appendix

\section{Evaluation of the Matsubara Frequency Summation}
\label{app:matsubara_evaluation}

In this Appendix, we provide the explicit evaluation of the polarization matrix elements $D_{\alpha\beta}(\bm{q}, i\omega_m)$. The unperturbed Green's function $\mathcal{G}_0(\bm{k}, i\omega_n)$ within the isolated flat-band subspace factorizes into a Nambu space component and a spatial Bloch state projector. Constrained by time-reversal symmetry, its spectral representation reads:
\begin{equation}
    \mathcal{G}_0(\bm{k}, i\omega_n) = \sum_{s \in \{3,4\}} \frac{|\phi_s\rangle\langle\phi_s|}{i\omega_n - E_s} \otimes |\psi_{\bm{k}}\rangle\langle\psi_{\bm{k}}|,
\end{equation}
where $E_3 = -E$ and $E_4 = E$ are the eigenvalues of the projected BdG matrix, $E = \sqrt{\mu^2 + \Delta_0^2}$, and $|\phi_s\rangle$ denotes the corresponding Nambu eigenvectors. 

The trace for the polarization bubble separates into a spatial and a Nambu component. The spatial trace yields the squared wave-function overlap $F(\bm{k}, \bm{q}) = \mathrm{Tr}_{\mathrm{space}}\left[ |\psi_{\bm{k}+\bm{q}}\rangle\langle\psi_{\bm{k}+\bm{q}}| \psi_{\bm{k}}\rangle\langle\psi_{\bm{k}}| \right] = \left| \langle \psi_{\bm{k}} | \psi_{\bm{k}+\bm{q}} \rangle \right|^2$. Consequently, the polarization matrix simplifies to:
\begin{equation} \label{eq:app_bubble_decomp}
    D_{\alpha\beta}(\bm{q}, i\omega_m) = \frac{1}{2} \sum_{\bm{k}} \left| \langle \psi_{\bm{k}} | \psi_{\bm{k}+\bm{q}} \rangle \right|^2 \sum_{s, s' \in \{3,4\}} t_\alpha^{s's} t_\beta^{ss'} \cdot \frac{1}{\beta} \sum_{i\omega_n} \frac{1}{(i\omega_n + i\omega_m - E_s)(i\omega_n - E_{s'})},
\end{equation}
where $t_\alpha^{s's} = \langle \phi_{s'} | \tau_\alpha | \phi_s \rangle$ are the Nambu matrix elements.

Standard contour integration of the internal fermionic Matsubara frequencies $\omega_n = (2n+1)\pi/\beta$ yields:
\begin{equation} \label{eq:matsubara_sum_formula}
    \frac{1}{\beta} \sum_{i\omega_n} \frac{1}{(i\omega_n + i\omega_m - E_s)(i\omega_n - E_{s'})} = \frac{n_F(E_{s'}) - n_F(E_s)}{i\omega_m + E_{s'} - E_s},
\end{equation}
where $n_F(x) = (e^{\beta x} + 1)^{-1}$ is the Fermi-Dirac distribution. 

At zero temperature ($T \rightarrow 0$), $n_F(E_3) = 1$ and $n_F(E_4) = 0$. The numerator $n_F(E_{s'}) - n_F(E_s)$ is non-vanishing only for interband transitions ($s \neq s'$), isolating the response into two distinct channels:
\begin{enumerate}
    \item Resonant channel ($s=4, s'=3$): $E_{s'} - E_s = -2E$ and $n_F(E_3) - n_F(E_4) = 1$. The propagator is $1/(i\omega_m - 2E)$.
    \item Anti-resonant channel ($s=3, s'=4$): $E_{s'} - E_s = 2E$ and $n_F(E_4) - n_F(E_3) = -1$. The propagator is $-1/(i\omega_m + 2E)$.
\end{enumerate}

To express these contributions compactly, we define:
\begin{align}
    T_1(\bm{k}, \bm{q}) &= t_\alpha^{34} t_\beta^{43} \left| \langle \psi_{\bm{k}} | \psi_{\bm{k}+\bm{q}} \rangle \right|^2, \\
    T_2(\bm{k}, \bm{q}) &= t_\alpha^{43} t_\beta^{34} \left| \langle \psi_{\bm{k}} | \psi_{\bm{k}+\bm{q}} \rangle \right|^2.
\end{align}
To evaluate these terms, we use the explicit Nambu eigenstates $|\phi_3\rangle = (-v, u)^T$ and $|\phi_4\rangle = (u, v)^T$, where the Bogoliubov coherence factors $u$ and $v$ are defined as:
\begin{equation}
    u = \sqrt{\frac{1}{2}\left(1 - \frac{\mu}{E}\right)}, \quad v = \sqrt{\frac{1}{2}\left(1 + \frac{\mu}{E}\right)}.
\end{equation}
Using these definitions, the transition matrix elements $t_\alpha^{34} = \langle \phi_3 | \tau_\alpha | \phi_4 \rangle$ evaluate directly to:
\begin{equation} \label{eq:app_nambu_elements}
    t_x^{34} = -\frac{\mu}{E}, \quad t_y^{34} = i, \quad t_z^{34} = -\frac{\Delta_0}{E}.
\end{equation}

The Hermiticity of the Pauli matrices ($\tau_\alpha^\dagger = \tau_\alpha$) ensures that $(t_\alpha^{34})^* = t_\alpha^{43}$, which gives:
\begin{equation} \label{eq:T1_T2_conjugate}
    T_2(\bm{k}, \bm{q}) = T_1^*(\bm{k}, \bm{q}).
\end{equation}

Substituting the non-vanishing channels and Eq.~(\ref{eq:T1_T2_conjugate}) into Eq.~(\ref{eq:app_bubble_decomp}), the polarization bubble is given by:
\begin{equation}
    D_{\alpha\beta}(\bm{q}, i\omega_m) = -\frac{1}{2} \sum_{\bm{k}} \left[ \frac{T_1(\bm{k}, \bm{q})}{2E - i\omega_m} + \frac{T_1^*(\bm{k}, \bm{q})}{2E + i\omega_m} \right].
\end{equation}
Performing the analytic continuation $i\omega_m \rightarrow \omega + i0^+$ yields the continuous real-frequency response function presented in Eq.~(\ref{eq:bubble_T1_T2}) of the main text.

\section{Derivation of the Zero-Frequency Pole}
\label{app:det_derivation}

To determine the zero-momentum frequency $\omega_0$ of the collective excitations in the long-wavelength limit ($q \to 0$), we solve the secular equation $\det[\bm{M}(\bm{q}=0, \omega_0)] = 0$. In the basis $(\rho, \Delta_0 \theta, \delta\phi)$, the response matrix $\bm{M}(0, \omega_0) = \bm{M}_{\text{bare}} + \bm{D}(0, \omega_0)$ is a $3 \times 3$ matrix. Evaluating the fermionic trace with the $T=0$ flat-band Green's functions and applying the mean-field gap equation $1/U = N_k/2E$ yields the matrix elements as functions of the chemical potential $\mu$, the pairing gap $\Delta_0$, and the energy $E = \sqrt{\mu^2 + \Delta_0^2}$:

\begin{align}
    M_{xx} &= \frac{N_k}{2E} \frac{4\Delta_0^2 - \omega_0^2}{4E^2 - \omega_0^2}, \\
    M_{yy} &= \frac{N_k}{2E} \frac{-\omega_0^2}{4E^2 - \omega_0^2}, \\
    M_{zz} &= \frac{N_k}{2E} \frac{4\mu^2 - \omega_0^2}{4E^2 - \omega_0^2}, \\
    M_{xz} = M_{zx} &= \frac{N_k}{2E} \frac{-4\mu \Delta_0}{4E^2 - \omega_0^2}, \\
    M_{xy} = -M_{yx} &= \frac{-i N_k \mu \omega_0}{E(4E^2 - \omega_0^2)}, \\
    M_{zy} = -M_{yz} &= \frac{-i N_k \Delta_0 \omega_0}{E(4E^2 - \omega_0^2)}.
\end{align}

For non-zero frequencies ($\omega_0 \neq 0$), the phase channel ($y$) dynamically couples with the amplitude ($x$) and density ($z$) channels through the off-diagonal terms $M_{xy}$ and $M_{zy}$, resulting in a fully hybridized $3 \times 3$ response matrix:
\begin{equation}
    \bm{M}(0, \omega_0) = \begin{pmatrix} 
    M_{xx} & M_{xy} & M_{xz} \\
    M_{yx} & M_{yy} & M_{yz} \\
    M_{zx} & M_{zy} & M_{zz}
    \end{pmatrix}.
\end{equation}

Evaluating the determinant of this matrix with the flat-band identity $E^2 = \mu^2 + \Delta_0^2$ yields:
\begin{equation}
    \det[\bm{M}(0, \omega_0)] = \frac{-N_k^3 \omega_0^2}{8E^3 (4E^2 - \omega_0^2)},
\end{equation}
where $N_k$ is the number of unit cells.

\section{Basis Vectors of the Null Space}
\label{app:null_space_derivation}

In this Appendix, we present the linearly independent basis vectors $|v_1\rangle$ and $|v_2\rangle$ that span the two-dimensional null space of the matrix $\bm{M}(0,0)$. Evaluated in the limit ($\bm{q} \to 0, \omega \to 0$) within the fluctuation basis $\Phi = (\rho, \Delta_0 \theta, \delta\phi)^T$, the matrix takes the form:
\begin{equation} \label{eq:app_M00_form}
    \bm{M}(0,0) = \frac{N_k}{2E^3} \begin{pmatrix} \Delta_0^2 & 0 & -\mu\Delta_0 \\ 0 & 0 & 0 \\ -\mu\Delta_0 & 0 & \mu^2 \end{pmatrix},
\end{equation}
where $\mu$ is the chemical potential, $\Delta_0$ is the uniform static pairing gap, and $E = \sqrt{\mu^2 + \Delta_0^2}$.

The matrix $\bm{M}(0,0)$ has rank 1, ensuring a two-dimensional null space. The first normalized basis vector corresponding to the zero eigenvalue aligns with the decoupled phase channel:
\begin{equation} \label{eq:app_v1_result}
    |v_1\rangle = \begin{pmatrix} 0 \\ 1 \\ 0 \end{pmatrix}.
\end{equation}

To construct the second orthogonal basis vector, we enforce the null space condition $\bm{M}(0,0)|v\rangle = \bm{0}$ for the remaining components, yielding the algebraic constraint $\Delta_0 v_x - \mu v_z = 0$. Normalizing this vector by the factor $E$ gives:
\begin{equation} \label{eq:app_v2_result}
    |v_2\rangle = \begin{pmatrix} \frac{\mu}{E} \\ 0 \\ \frac{\Delta_0}{E} \end{pmatrix}.
\end{equation}

\section{Reduction to the Effective Matrix}
\label{app:schur_justification}

In the main text, the dispersion relation of the collective modes is determined by the roots of the secular equation $\det[\bm{M}(\bm{q}, \omega)] = 0$. To justify the dimension reduction of this $3 \times 3$ response matrix, we utilize the properties of block matrices.

We apply an orthogonal basis transformation to partition the full response matrix $\bm{M}$ into the low-energy null space $\mathcal{P}$ (denoted by index $L$) and its orthogonal complement $\mathcal{Q}$ (denoted by index $H$). The matrix takes the block form:
\begin{equation}
    \bm{M} = \begin{pmatrix} \bm{M}_{LL} & \bm{M}_{LH} \\ \bm{M}_{HL} & \bm{M}_{HH} \end{pmatrix},
\end{equation}
where $\bm{M}_{LL}$ represents the projection of the matrix $\bm{M}$ onto the null space.

According to the Schur determinant identity, the determinant of this partitioned matrix can be factored as:
\begin{equation} \label{eq:schur_determinant_identity}
    \det(\bm{M}) = \det(\bm{M}_{HH}) \cdot \det(\bm{M}_{LL} - \bm{M}_{LH} \bm{M}_{HH}^{-1} \bm{M}_{HL}).
\end{equation}

In the long-wavelength and low-energy limit, the block $\bm{M}_{HH}$ is non-singular ($\det \bm{M}_{HH} \neq 0$). Consequently, the roots of the full secular equation are identical to the roots of the reduced effective equation:
\begin{equation}
    \det(\bm{M}_{\text{eff}}) = 0, \quad \text{where} \quad \bm{M}_{\text{eff}} = \bm{M}_{LL} - \bm{M}_{LH} \bm{M}_{HH}^{-1} \bm{M}_{HL}.
\end{equation}

In our perturbation expansion with respect to the momentum $q$, the effective matrix $\bm{M}_{\text{eff}}$ is evaluated order by order. As demonstrated in the main text, the Schur complement correction term $\bm{M}_{LH} \bm{M}_{HH}^{-1} \bm{M}_{HL}$ constitutes a higher-order correction compared to the leading non-vanishing terms in $\bm{M}_{LL}$ at specific perturbation orders. 

When this higher-order correction term is neglected at leading order, the effective matrix simplifies to the direct null-space projection: $\bm{M}_{\text{eff}} \approx \bm{M}_{LL}$. Therefore, solving the full secular equation $\det(\bm{M}) = 0$ is equivalent to solving the reduced projected equation $\det(\bm{M}_{LL}) = 0$ within the controlled precision of the perturbation expansion.

\section{Derivation of Matrix \texorpdfstring{$\bm{A}$}{A}}
\label{app:matrix_A_evaluation}

\subsection{First-Order Frequency Derivative of the Response Matrix}
Since the bare interaction matrix $\bm{M}_{\text{bare}}$ is constant, the frequency dependence of the response matrix $\bm{M}(\bm{q}, \omega)$ arises entirely from the polarization bubble given in Eq.~(\ref{eq:bubble_T1_T2}). Taking the partial derivative with respect to $\omega$ and evaluating it at the static origin $(\bm{q}=\bm{0}, \omega=0)$ yields:
\begin{equation} \label{eq:app_derivative_general}
    \left. \frac{\partial \bm{M}_{\alpha\beta}}{\partial \omega} \right|_{(0,0)} = \left. \frac{\partial D_{\alpha\beta}}{\partial \omega} \right|_{(0,0)} = -\frac{1}{8E^2} \sum_{\bm{k}} \left[ T_1(\bm{k}, 0) - T_2(\bm{k}, 0) \right].
\end{equation}

At $\bm{q} = \bm{0}$, the spatial wave-function overlap evaluates to unity ($|\langle \psi_{\bm{k}} | \psi_{\bm{k}} \rangle|^2 = 1$). Consequently, the terms $T_1$ and $T_2$ reduce to the products of Nambu matrix elements: $T_1(\bm{k}, 0) = t_\alpha^{34} t_\beta^{43}$ and $T_2(\bm{k}, 0) = t_\alpha^{43} t_\beta^{34}$. Using the matrix elements established in Eq.~(\ref{eq:app_nambu_elements}), the diagonal channels ($\alpha = \beta$) vanish identically. The non-vanishing cross-coupled channels evaluate to:
\begin{equation}
    T_1(xy) - T_2(xy) = 2i \frac{\mu}{E}, \quad T_1(zy) - T_2(zy) = 2i \frac{\Delta_0}{E}.
\end{equation}

Since the terms are independent of $\bm{k}$, summing over the Brillouin zone directly yields the number of unit cells $N_k$. Enforcing $\partial_\omega M_{\alpha\beta} = -\partial_\omega M_{\beta\alpha}$, we obtain the explicit matrix representation:
\begin{equation} \label{eq:app_dM_dw_explicit_proof}
    \left. \frac{\partial \bm{M}}{\partial \omega} \right|_{(0,0)} = \frac{N_k}{4E^3} \begin{pmatrix} 0 & -i\mu & 0 \\ i\mu & 0 & i\Delta_0 \\ 0 & -i\Delta_0 & 0 \end{pmatrix}.
\end{equation}

\subsection{Projection onto the Low-Energy Subspace}
To construct the effective matrix $\bm{A}$, we project the matrix derived in Eq.~(\ref{eq:app_dM_dw_explicit_proof}) onto the two-dimensional null space using $\bm{A}_{ij} = \langle v_i | \left. \frac{\partial \bm{M}}{\partial \omega} \right|_{(0,0)} | v_j \rangle$. Recalling the basis vectors from Appendix~\ref{app:null_space_derivation}:
\begin{equation}
    |v_1\rangle = \begin{pmatrix} 0 \\ 1 \\ 0 \end{pmatrix}, \quad |v_2\rangle = \begin{pmatrix} \frac{\mu}{E} \\ 0 \\ \frac{\Delta_0}{E} \end{pmatrix},
\end{equation}
the projection directly yields $\bm{A}_{11} = \bm{A}_{22} = 0$ for the diagonal elements. For the off-diagonal elements, the inner product gives:
\begin{align} \label{eq:app_A_elements_direct}
    \bm{A}_{12} &= \frac{N_k}{4E^3} \left( i\mu \frac{\mu}{E} + i\Delta_0 \frac{\Delta_0}{E} \right) = i\frac{N_k (\mu^2 + \Delta_0^2)}{4E^4} = i\frac{N_k}{4E^2}, \\
    \bm{A}_{21} &= \frac{N_k}{4E^3} \left( -i\mu \frac{\mu}{E} - i\Delta_0 \frac{\Delta_0}{E} \right) = -i\frac{N_k (\mu^2 + \Delta_0^2)}{4E^4} = -i\frac{N_k}{4E^2}.
\end{align}

Therefore, the projected effective matrix $\bm{A}$ takes the form:
\begin{equation} \label{eq:app_A_final_canonical}
    \bm{A} = \frac{N_k}{4E^2} \begin{pmatrix} 0 & i \\ -i & 0 \end{pmatrix}.
\end{equation}

\section{Derivation of Matrix \texorpdfstring{$\bm{B}_{ab}$}{B\_ab}}
\label{app:matrix_B_evaluation}

\subsection{Second-Order Momentum Expansion of the Response Matrix}
Since the bare interaction matrix $\bm{M}_{\text{bare}}$ is momentum-independent, the momentum derivatives of the response matrix at $(\bm{q}=\bm{0}, \omega=0)$ arise entirely from the polarization bubble $\bm{D}(\bm{q}, \omega)$. Applying the differential operator $\frac{\partial^2}{\partial q_a \partial q_b}$ to Eq.~(\ref{eq:bubble_T1_T2}) yields:
\begin{equation} \label{eq:app_bubble_second_derivative_step1}
    \left. \frac{\partial^2 \bm{M}_{\alpha\beta}}{\partial q_a \partial q_b} \right|_{(0,0)} = -\frac{1}{4E} \sum_{\bm{k}} \left. \frac{\partial^2}{\partial q_a \partial q_b} \left[ T_1(\bm{k}, \bm{q}) + T_2(\bm{k}, \bm{q}) \right] \right|_{\bm{q}=\bm{0}}.
\end{equation}

Using the relation $T_2(\bm{k}, \bm{q}) = T_1^*(\bm{k}, \bm{q})$, we have $T_1 + T_2 = 2 \mathrm{Re}[T_1]$. Recalling the explicit form $T_1(\bm{k}, \bm{q}) = t_\alpha^{34} t_\beta^{43} |\langle \psi_{\bm{k}} | \psi_{\bm{k}+\bm{q}} \rangle|^2$ and noting that $t_\alpha^{34}$ and $t_\beta^{43}$ are independent of $\bm{q}$, the momentum derivatives act exclusively on the wave-function overlap. 
Using the small-momentum expansion $|\langle \psi_{\bm{k}} | \psi_{\bm{k}+\bm{q}} \rangle|^2 = 1 - \sum_{i,j} q_i q_j g_{ij}(\bm{k}) + \mathcal{O}(q^3)$, where $g_{ij}(\bm{k})$ is the quantum metric tensor, the second-order derivative evaluates to:
\begin{equation} \label{eq:app_overlap_second_derivative}
    \left. \frac{\partial^2}{\partial q_a \partial q_b} \left| \langle \psi_{\bm{k}} | \psi_{\bm{k}+\bm{q}} \rangle \right|^2 \right|_{\bm{q}=\bm{0}} = -2 g_{ab}(\bm{k}).
\end{equation}

Substituting Eq.~(\ref{eq:app_overlap_second_derivative}) into Eq.~(\ref{eq:app_bubble_second_derivative_step1}), we obtain:
\begin{equation} \label{eq:app_half_second_derivative_final}
    \frac{1}{2} \left. \frac{\partial^2 \bm{M}_{\alpha\beta}}{\partial q_a \partial q_b} \right|_{(0,0)} = \frac{1}{2E} \sum_{\bm{k}} g_{ab}(\bm{k}) \mathrm{Re}\left[ t_\alpha^{34} t_\beta^{43} \right].
\end{equation}

\subsection{Subspace Projection and Identity Matrix Representation}
The effective matrices $\bm{B}_{ab}$ are obtained by projecting Eq.~(\ref{eq:app_half_second_derivative_final}) onto the null space: $\left[\bm{B}_{ab}\right]_{ij} = \frac{1}{2} \langle v_i | \left. \frac{\partial^2 \bm{M}}{\partial q_a \partial q_b} \right|_{(0,0)} | v_j \rangle$. Using the explicit matrix elements from Eq.~(\ref{eq:app_nambu_elements}), the real part $\mathrm{Re}\left( t_\alpha^{34} t_\beta^{43} \right)$ evaluates to the following:
\begin{equation} \label{eq:app_Re_t_matrix}
    \left[ \mathrm{Re}\left( t_\alpha^{34} t_\beta^{43} \right) \right]_{\alpha\beta} = \begin{pmatrix} \frac{\mu^2}{E^2} & 0 & \frac{\mu\Delta_0}{E^2} \\ 0 & 1 & 0 \\ \frac{\mu\Delta_0}{E^2} & 0 & \frac{\Delta_0^2}{E^2} \end{pmatrix}.
\end{equation}

Projecting this matrix onto the basis vectors $|v_1\rangle = (0, 1, 0)^T$ and $|v_2\rangle = (\mu/E, 0, \Delta_0/E)^T$, the diagonal elements ($i=j$) yield:
\begin{align}
    \langle v_1 | \left[ \mathrm{Re}\left( t_\alpha^{34} t_\beta^{43} \right) \right] | v_1\rangle &= 1, \\
    \langle v_2 | \left[ \mathrm{Re}\left( t_\alpha^{34} t_\beta^{43} \right) \right] | v_2\rangle &= \frac{\mu^4}{E^4} + 2\frac{\mu^2\Delta_0^2}{E^4} + \frac{\Delta_0^4}{E^4} = \frac{(\mu^2 + \Delta_0^2)^2}{E^4} = 1,
\end{align}
where the identity $\mu^2 + \Delta_0^2 = E^2$ is used. The off-diagonal terms ($i \neq j$) vanish ($\left[\bm{B}_{ab}\right]_{12} = \left[\bm{B}_{ab}\right]_{21} = 0$).

Consequently, the effective matrices $\bm{B}_{ab}$ are proportional to the $2 \times 2$ identity matrix $\bm{I}$:
\begin{equation} \label{eq:app_B_final_diagonal}
    \bm{B}_{ab} = \frac{1}{2E} \sum_{\bm{k}} g_{ab}(\bm{k}) \begin{pmatrix} 1 & 0 \\ 0 & 1 \end{pmatrix} \equiv \bar{B}_{ab} \bm{I}.
\end{equation}

\end{document}